\documentclass[amsmath,amssymb,twocolumn]{revtex4}
\usepackage[parfill]{parskip}    
\usepackage{graphicx}
\usepackage{amssymb}
\usepackage{epstopdf}
\usepackage{color}
\usepackage{graphicx}
\usepackage{pdfpages}
\usepackage{hyperref}

\begin{document}
\title{Relativistic dust accretion onto a scale--dependent polytropic black hole}
\author{Ernesto Contreras $^1$\footnote{On leave from Universidad Central de Venezuela}\footnote{econtreras@yachaytech.edu.ec}
\'Angel Rinc\'on $^2$\footnote{arrincon@uc.cl}
and J. M., Ram{\'i}rez-Velasquez $^1$\footnote{josem.ramirez@gmail.com, jmramirez@yachaytech.edu.ec}
}
\affiliation{
$^1$Yachay Tech, School of Physical Sciences \& Nanotechnology, 
100119, Urcuqu{\'i}, Ecuador\\
$^2$Instituto de F{\'i}sica, Pontificia Universidad Cat{\'o}lica de Chile, Av. Vicu{\~n}a Mackenna 4860, Santiago, Chile 
}
\begin{abstract}
In this work we study steady and spherical relativistic dust accretion onto a static and spherically symmetric scale--dependent black hole. In particular we consider a polytropic scale--dependent black hole as a central object and obtain that the radial velocity profile and the energy density are affected when scale--dependence of the central object is taken into account and such a deviation is controlled by the so called running parameters of the scale--dependence models.

\end{abstract}

\maketitle

\section{Introduction}
\label{intro}
Accretion of matter is one of the most important phenomena in the astrophysical realm. In fact studies on X-ray binaries, active galactic nuclei (AGNs), tidal disruption events, and gamma-ray bursts are based on accretion processes.
The first studies on accretion of matter were considered in the context of Newtonian gravity \cite{hoyle1939,bondi1944,bondi1952} and then generalized to curved space--times in \cite{michael1972}. 
%
Recently analytical work on isothermal Bondi-like accretion including radiation pressure and the gravitational potential of the host galaxy has risen enthusiasm on the galatic evolution theory community \citep{korol16a,ciotti17a,ciotti18a}. Moreover, detailed numerical computations on Bondi accretion \citep{ramirez18a} using novel consistent SPH (Smoothed Particle Hydrodynamics) techniques \citep{gabbasov17a,sigalotti18a} promises to push even further studies at sub-parsec scales in AGNs, even including radiation pressure due to lines \citep{ramirez16a,ramirez16b}.
Besides, accretion process have been consider in the context of 
General Relativity and models beyond the classical Einstein field equations with different 
interests \cite{belgman1978,pretrich1988,malec1999,babichev2004,babichev2005,karkowski2006,mach2008,gao2008,
jamil 2008,giddings2008,jimenez2008,sharif2011,dokuchaev2011,bibichev2012,bhandra2012,mach2013,mach2013a,
karkowski2013,jhon2013,ganguly2014,bachivev2014,debnath2015,yang2015,bahamonde2017,jiao2017,
jiao2017a,paik2018}. More recently, quantum correction to general relativistic accretion have been considered in Ref. \cite{yang2015}. 

In this work we study accretion to test scale--dependent models which are inspired in the well known asymptotic safety program 
\cite{weinberg1979,Wetterich:1992yh,Dou:1997fg,Souma:1999at,Reuter:2001ag,Fischer:2006fz,Percacci:2007sz,Litim:2008tt}. 
Scale--dependent gravitational theories
have been extensively used to obtain modified solution of the Einstein field equations  in  three dimensional space--times 
\cite{koch2016,
Rincon:2017ypd,
Rincon:2017goj,
rincon2018,
Rincon2018b}, 
four dimensional black holes 
\cite{Rincon:2017ayr,
Rincon2018,
koch2015a,
koch2015,
koch2015b,
contreras2018,
Contreras2018,
contreras2018c},
cosmological models \cite{Koch:2010nn,Hernandez2018} and traversable wormhole solutions 
\cite{contreras2018b}. One of the most interesting aspects of scale--dependent models 
is the apparition of some running parameter which controls the deviations form the classical Einstein General Relativity. Among the most interesting results obtained with scale--dependent gravity are modifications in the horizon
radius, asymptotic behavior and black hole thermodynamics. The above mentioned deviations are thought to be important in situations where the classical General Relativity is not longer valid. The study of accretion onto scale--dependent black holes
could serve as an useful tool in to confirm the validity of those models. In this sense, we are interesting in knowing how the accretion process is modified
when the central object is slightly deviated from the classical one. 
In this paper, as a continuation of a previous work \cite{Contreras2018}, 
we consider a scale--dependent polytropic black hole but this time as a central object 
responsible of the accretion of dust. As it is well known, the classical (non--scale--dependent) polytropic black hole \cite{setare}
is a novelty solution obtained after mapping the negative cosmological coupling with an effective pressure and demanding that it obeys a polytropic equation of state. After that, the matter content degrees of freedom are eliminated from the Einstein field equations and, finally, solutions matching polytropic thermodynamics with that of black holes are obtained. The matter sector arising from this protocol results in an anisotropic matter with the attractive classical feature that it fulfill all the energy conditions. More recently, we obtained that the introduction of scale--dependence in the classical polytropic solution leads to modifications in the black hole thermodynamics and changes in the topology of the space--time \cite{contreras2018}. In this sense, the polytropic black hole solution and its scale--dependent counterpart represent an interesting system to be taken into account. Even more, as the scale--dependence geometry contains the classical case, the main goal of this paper is to study accretion onto the classical polytropic solution and to compare it with its scale--dependent  case.\\
This work is organized as follows. Section \ref{general} is devoted to summarize the main aspects of spherically symmetric accretion. 
In Sect. \ref{scales} we review some aspects related to scale--dependent
gravity. In section \ref{accretion} we show our results and final comments are left to the concluding remarks on 
Sect. \ref{remarks}

\section{Accretion process of general static spherically symmetric black hole}\label{general}
We consider the following metric ansatz for the general
static spherically symmetric space--time
\begin{eqnarray}
\mathrm{d}s^{2} = -f(r) \mathrm{d} t^{2}+ f(r)^{-1} \mathrm{d}r^{2}+r^{2}\mathrm{d}\Omega^{2},
\end{eqnarray}
where $A(r)>0$ is a functions of $r$ only.

The energy-momentum tensor for the fluid is given by
\begin{eqnarray}
T_{\mu\nu}=(\rho+p)u_{\mu}u_{\nu}+g_{\mu\nu}p,
\end{eqnarray}
where $\rho$, $p$ $u^{\mu}$ are the energy density, the pressure and the four velocity
of the fluid.

The basic equations for the fluid are, the conservation of mass flux
\begin{eqnarray}
\nabla_{i}J^{i}=0,
\end{eqnarray}
and the energy flux
\begin{eqnarray}
\nabla_{i}T^{i}_{0}.
\end{eqnarray}
The above equations can be simplified in the case of steady state conditions and spherical symmetry 
giving
\begin{eqnarray}\label{cons1}
\frac{\mathrm{d}}{\mathrm{d}r}(J^{1}r^{2})=0,
\end{eqnarray}
and
\begin{eqnarray}\label{cons2}
\frac{\mathrm{d}}{\mathrm{d}r}(T^{1}_{0}r^{2})=0.
\end{eqnarray}
Integration of equations \eqref{cons1} and \eqref{cons2} leads to
\begin{eqnarray}
\rho r^{2}u&=&C_{1}\label{uno},\\
(p+\rho)u_{0} u r^{2}&=&C_{2}\label{dos},
\end{eqnarray}
where $C_{1}$ and $C_{2}$ are integration constants and $u^{0}$ and $u=u^{1}$ are non-zero
components of the velocity vector satisfying $g_{00}u^{0}u^{0}+g_{11}u^{1}u^{1}=-1$. Combining
\eqref{uno} and \eqref{dos} we obtain, 
\begin{eqnarray}
\frac{(p+\rho)^{2}}{\rho^{2}}(f+u^{2})=\left(\frac{C_{2}}{C_{1}}\right)^{2}=C_{3}\label{tres}.
\end{eqnarray}
Note that given an equation of state that relates $p$ and $\rho$, we have two equations
and two unknowns $\rho$ and $u$. 

The above equations are characterized by a critical point, as is usual for hydrodynamic
flow systems. Differentiation of \eqref{uno} and \eqref{tres} and elimination of $d\rho$ lead to
\begin{eqnarray}\label{criticalcond}
\frac{u'}{r}\left[V^{2}-\frac{u^{2}}{f^{2}+u^{2}}\right]
+\frac{1}{r}\left[2V^{2}-\frac{r f'}{2(f^{2}+u^{2})}\right]=0.
\end{eqnarray}
It is evident that if one or the other of the bracketted factors in \eqref{criticalcond}
vanishes one has turn-around point, and the solutions are double-valued in either $r$ or $u$. Only solutions that pass through a critical point correspond to material falling into (or flowing
out of) the object with monotonically increasing velocity along the particle trajectory.
The critical point is located where both bracketted factors in Eq. \eqref{criticalcond} vanish, thus
\begin{eqnarray}
u_{c}^{2}&=&\frac{r f_{c}'}{4},\\
V_{c}^{2}&=&\frac{u^{2}_{c}}{f_{c}^{2}+u_{c}^{2}}.
\end{eqnarray}
\section{Scale--dependent polytropic black hole}
\label{scales}
In this section we shall explore the main results obtained in the context of scale--dependent gravity following references 
\cite{koch2016,Rincon:2017ypd,Rincon2018b,Rincon:2017goj,rincon2018,Rincon:2017ayr,Rincon2018,koch2015a,koch2015,koch2015b,contreras2018,Contreras2018}
%
The effective Einstein--Hilbert action considered here reads
\begin{eqnarray}\label{action}
S[g_{\mu\nu},k]=\int \mathrm{d}^{4}x \sqrt{-g} \Bigg[\frac{1}{2 \kappa_k} R + \mathcal{L}_{M} \Bigg],
\end{eqnarray}
with $\kappa_k \equiv 8 \pi G_{k}$ being the Einstein coupling, $G_{k}$ standing for the scale--dependent gravitational coupling, and $\mathcal{L}_{M}$ is the Lagrangian density which correspond to the matter sector. The scale--dependent action provides: i) the effective Einstein field equation (when we vary respect the metric field), and ii) a self--consistent equation (when we vary respect the scalar field $k$). What is more, $k$ is usually connected which a energy scale and encoded any possible quantum effect, if it is present. Then, the effective Einstein equations are
%
\begin{eqnarray}\label{einstein}
G_{\mu\nu} = \kappa_k T^{eff}_{\mu\nu},
\end{eqnarray}
where $T^{eff}_{\mu\nu}$ is the effective energy momentum tensor defined according to
\begin{eqnarray}\label{eff}
\kappa_{k} T^{eff}_{\mu\nu}:= \kappa_{k} T_{\mu\nu} - \Delta t_{\mu\nu},
\end{eqnarray}
$T_{\mu\nu}$ corresponds to the matter energy--momentum tensor and $\Delta t_{\mu\nu}$, given by
\begin{eqnarray}\label{nme}
\Delta t_{\mu\nu} = G_{k} \Bigl( g_{\mu\nu}\square -\nabla_{\mu}\nabla_{\nu} \Bigl) G_{k}^{-1},
\end{eqnarray}
is the so--called non--matter energy--momentum tensor. Thus, the above tensor parametrize the inclusion of any quantum effect via the running of the gravitational coupling. Notice that the running gravitational coupling does not have dynamics which is a important difference between our approach and the Brans-Dicke scenario \cite{Brans:1961sx}.
The corresponding variation of the effective action, with respect to the scale--field $k(x)$, provides an auxiliary equation
\begin{eqnarray}\label{scale}
\frac{\delta S[g_{\mu\nu},k]}{\delta k}=0.
\end{eqnarray}
In principle, if we combine Eq. \eqref{einstein} with the obtained from Eq. \eqref{scale}, the fields
involved might be determined. In particular, the scale setting equation \eqref{scale} allows us
to determine the scalar function $k(x)$. However, at least some functional form of $G_{k}$ is given by certain beta function for example, the problem remains unsolved. 
In order to elude the aforementioned difficulty, we can reasoning as follow: we know that $G_{k}$ inherit some dependence of the coordinates from $k(x)$ and therefore one might treat this as an independent field $G(x)$. Thus, in what follow, we will treat directly the couplings as functions of the radial coordinate, $(\cdots)(r)$, instead of the energy, $(\cdots)_k$. 

For the purpose of the present work, we consider a static and spherically symmetric space--time with a line element
 parametrized 
as
\begin{eqnarray}\label{metricp}
\mathrm{d}s^2 = -f(r) \mathrm{d} t^2 + f(r)^{-1} \mathrm{d}r^2 + r^2 \mathrm{d} \Omega^2. 
\end{eqnarray}

It is worth to noticing that after replacing Eq. (\ref{metricp}) in Eq. (\ref{einstein}), three independent 
differential equations for the four independent fields $f(r)$, $G(r)$, $T^{0}_{0}$ and $T^{2}_{2}$ are obtained. An alternative way to decrease
the number of degrees of freedom consists in demanding
some energy condition on $T_{eff}$. In this work we adopt the same strategy as in  and we demand the null energy condition (NEC)
for being the least restrictive condition we can employ to obtain suitable solutions.
More precisely, for the effective energy momentum tensor, the NEC reads
\begin{eqnarray}
\kappa(r)
T^{eff}_{\mu\nu}n^{\mu}n^{\nu}:= \kappa(r)T_{\mu\nu}n^{\mu}n^{\nu} - \Delta t_{\mu\nu}n^{\mu}n^{\nu},
\end{eqnarray}
where $n^{\mu}$ is a null vector. With the parametrization of Eq. (\ref{metricp}), both $G_{\mu\nu}$ and $T_{\mu\nu}$ 
saturate the NEC and, therefore, 
\begin{eqnarray}
\Delta t_{\mu\nu}n^{\mu}n^{\nu}=0
\end{eqnarray}
for consistency. The above condition leads to a differential equation for $G(r)$ given by
\begin{eqnarray}
2\left(\frac{ \mathrm{d} G}{\mathrm{d}r}\right)^2 -G \frac{\mathrm{d}^{2}G}{\mathrm{d} r^{2}}=0,
\end{eqnarray}
from where
\begin{eqnarray}\label{gr}
G(r)=\frac{G_{0}}{1+\epsilon r},
\end{eqnarray}
with $\epsilon\ge 0$ is a parameter with 
dimensions of inverse of length. It is worth mentioning that, in the limit 
$\epsilon\to 0$, $G(r)=G_{0}$, $\Delta t_{\mu\nu}=0$ 
and the classical Einstein's field equations are recovered. For this reason, $\epsilon$ is called the running parameter, which
controls the strength of the scale--dependency. 
The solution for the scale--dependent 
polytropic black \cite{Contreras2018} hole is given by
\begin{equation}
\begin{split}
&f(r) = f_{0}(r) + 6 G_{0} M_{0} r^2 \epsilon ^3 \ln \left[2 G_{0} M_{0} \frac{r \epsilon +1}{r}\right] \\
& \hspace{1cm} +3G_{0}M_{0}\epsilon  (1-2 r \epsilon ), \label{lapsepoly}
\end{split},
\end{equation}
where $M_{0}$ corresponds to the classical BH mass and
\begin{equation} \label{f_class}
f_{0}(r)= \bigg(\frac{r}{L}\bigg)^2 \bigg[1 - \bigg( \frac{r_{0}}{r} \bigg)^3 \bigg],
\end{equation}
stands for the classical polytropic BH solution (without running) where
\begin{equation}
r_{0}=\sqrt[3]{2 G_0 L^2 M_0}.
\end{equation}

Note that for $\epsilon \ll 1$ Eq. (\ref{lapsepoly}) takes the simply form
\begin{equation}\label{fra}
f(r)\approx  3 G_0 M_0 \epsilon-\frac{2 G_0 M_0}{r} +\frac{r^2}{L^2},
\end{equation}
or, in term of the classical parameters we can write down the lapse function as
\begin{align} \label{f_style}
f(r) &\approx \bigg(\frac{r}{L}\bigg)^2 \bigg[1 - a \bigg( \frac{r_{0}}{r} \bigg)^3 \bigg],
\end{align}
where the auxiliary function $a \equiv 1- (3/2) \epsilon r$. The aforementioned relation means that the scale--dependent effect only alter a concrete sector of the solution and, of course, \ref{f_style} converge to \ref{f_class} when $\epsilon$ is taken to be zero.
It is worth noticing that the above metric function (see Eq. (\ref{fra})) corresponds to a $4$-dimensional Schwarzschild-Anti de Sitter black hole in the presence of an external string cloud \cite{dey2017} with the string cloud parameter given by $\alpha=3M_{0}G_{0}\epsilon-1$.
In this work, we are interested in to obtain small deviations in the accretion process respect to the expected classical results. Therefore, despite the exact solution of the scale--dependent polytropic black hole is known, we will focus our attention on the case where the running parameter is considered small, compared with the other relevant scales in the problem. The reason is that the scale--dependent philosophy assume that any quantum correction should be small and, as $\epsilon$ control the strength of the gravitational coupling, we finally assume small values of that parameter. We then take advantage of this fact to make progress. For the aforementioned reason, 
we will study the accretion onto the scale--background described by the approximated metric in Eq. (\ref{fra}). 

\section{Relativistic dust accretion}\label{accretion}
In order to describe the accretion process we must be able to obtain the radial velocity profile $u(r)$ and the density $\rho$ from Eqs. (\ref{uno}) and (\ref{tres}). Of course, depending on the nature of the matter content, the pressure can be obtained as a function of $\rho$ from the equation of the state of the accreted matter. In the case of dust we set $p=0$ so that  Eq. (\ref{tres}) can be trivially decoupled \footnote{Note that the equation can be trivially  decoupled whenever $p=\alpha \rho$ with $\alpha=$constant} and can be written as
\begin{align}
u(r) & = - \ \sqrt{\left(\frac{C_1}{C_2}\right)^2 - f(r) }
\end{align}
For the particular metric in Eq. (\ref{fra}), the velocity profile reads
\begin{eqnarray}\label{ur}
u(r)=-\frac{\sqrt{C_1^2 L^2 r-C_2^2 \left(G_0 L^2 M_0 (3 r \epsilon -2)+r^3\right)}}{C_2 L \sqrt{r}}.
\end{eqnarray}

Replacing (\ref{ur}) in (\ref{uno}) the density profile is given by
\begin{align}
\rho(r) &=-\frac{C_{1} C_2 L \sqrt{r}}{r^2\sqrt{C_1^2 L^2 r-C_2^2 \left(G_0 L^2 M_0 (3 r \epsilon -2)+r^3\right)}}
\end{align}
Note that in order to obtain ingoing fluid and a positive energy density we demand $C_{1} < 0$ and $C_{2}>0$.

\begin{figure}
\centering
\includegraphics[width=\linewidth]{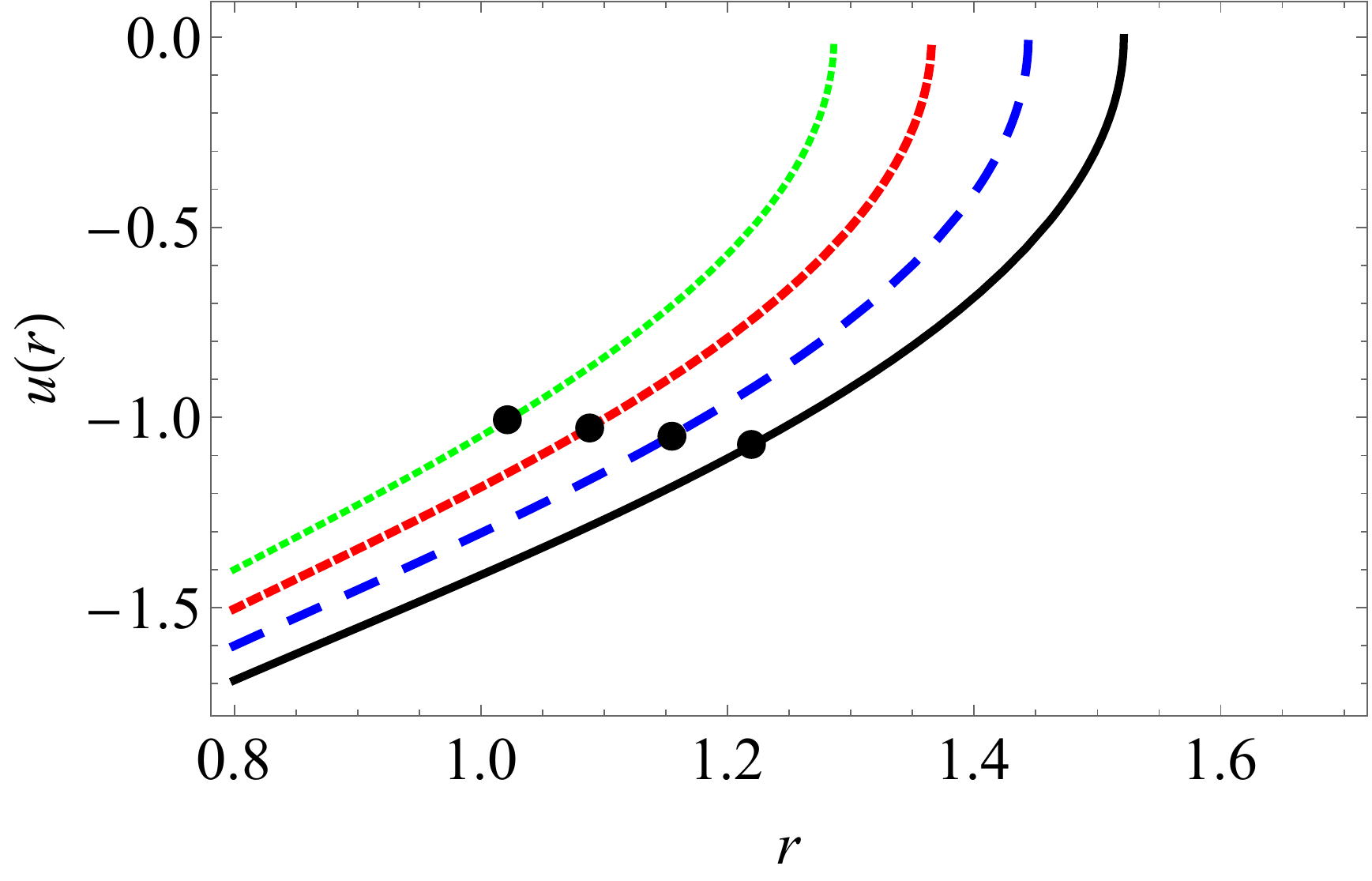}
\caption{\label{lapsefig} 
Radial velocity for $C_{1}=-1$, $C_{2}=1$
$\epsilon=0.00$ (black solid line), $\epsilon=0.1$ (dashed blue line), $\epsilon=0.20$ 
(short dashed red line) and $\epsilon=0.30$ (dotted green line). The other values have been taken as unity. The dots depicts the critical points of each solution. See text for details.
}
\end{figure}

In Fig. \ref{lapsefig} it is shown the velocity profile for different values of the running parameter $\epsilon$. The dots denote the critical velocity of the fluid at a certain critical radius. Note that when $\epsilon$ increase the critical point
shift to the left. It is worth noticing that this shift in the critical point appears in other contexts. For example, in Ref. \cite{bahamonde2017} the critical point undergoes a shift but in this case it is due to the nature of the accreted 
matter, {\it i.e}, for variations in the parameter $\omega$ of the equation of state $p=\omega \rho$ in Eq. (\ref{tres}). In this case it was obtained that as $\omega$ increase the critical point moves towards decreasing radius. In this sense, the behavior of the radial velocity if either the central object change (scale--dependent model) or if the accreted content varies (fixed central object in Ref. \cite{bahamonde2017}) is formally the same.

\begin{figure}
\centering
\includegraphics[width=\linewidth]{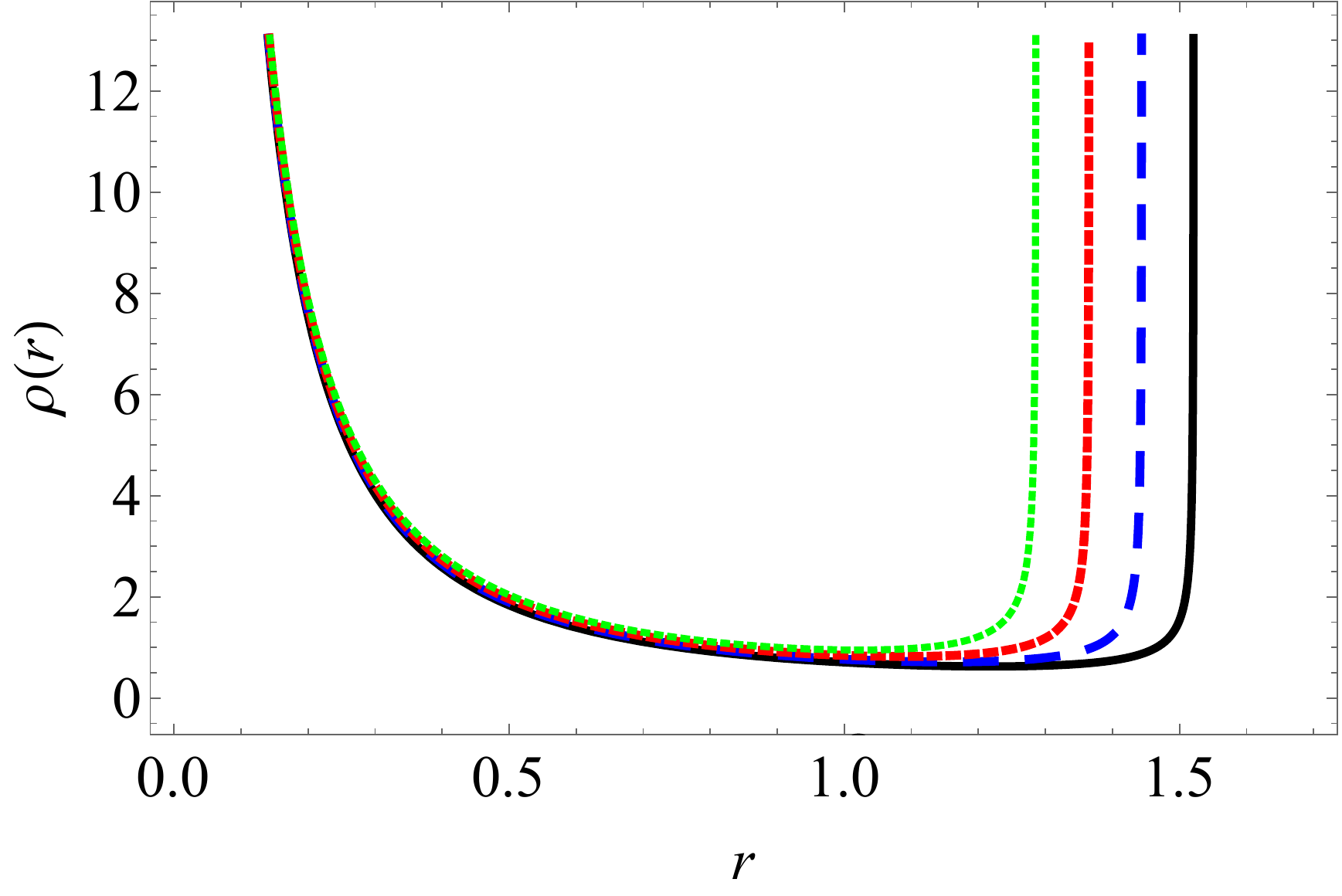}
\caption{\label{rhofig} 
Energy density for for $C_{1}=-1$, $C_{2}=1$
$\epsilon=0.00$ (black solid line), $\epsilon=0.1$ (dashed blue line), $\epsilon=0.20$ 
(short dashed red line) and $\epsilon=0.30$ (dotted green line). The other values have been taken as unity. See text for details.
}
\end{figure}

In figure \ref{rhofig} we show the
energy density profile. 
Note that, on one hand the energy density increases as the fluid moves towards the black hole. On the other hand, the effect of the scale can be appreciated far from the black hole because
in its vicinity the behaviour is indistinguishable. It is worth noticing that the behaviour depicted in Fig. \ref{rhofig} coincide with
that reported in Ref. \cite{bahamonde2017} for
a Schwarzschild Black Hole in a string cloud. Indeed, this is an expected result because, as commented before, first order corrections in $\epsilon$ of the metric function (see Eq. (\ref{fra})) leads to a Schwarzschild-Anti de Sitter black hole in the presence of external string cloud.
\section{Concluding remarks}
\label{remarks}
In this work, we have considered for the first time the accretion process onto a scale--dependent space--time.
%
It is remarkable the formal similarity between the results obtained here and those obtained in other contexts (see Ref. \cite{bahamonde2017}, for example). 
Should be notice that the scale--dependent framework introduce certain deviations respect the classical counterpart and, indeed, the velocity profile as well as the energy density profile are now lower than the standard solution.
In fact, the main difference of our findings respect to other approaches is that in this work we studied the effects of scale--dependence considering a fixed matter content instead of fixing the background and varying the accreted matter. To be more precise, we modified the central object through the running parameter instead of tuning the parameter of the equation of state that, on the contrary, modify the nature of the accreted fluid. It is worth noticing that the simple model consider here could shed some lights about how the scale--dependence modify the accretion process. It could be interesting testing the scale--dependent effects in a more realistic scenario however this subject goes far beyond the scope of this work and will be worked out in a future work.

\end{document}